\documentclass[prc,twocolumn,showpacs,preprintnumbers,amsmath,amssymb]{revtex4}
\usepackage{rotating}
\usepackage{graphicx}
\usepackage{dcolumn}
\usepackage{amstext}
\usepackage{lscape}
\usepackage{verbatim}
\usepackage[colorlinks,citecolor=blue,bookmarks]{hyperref}
\usepackage{times}
\usepackage{bm} 
\begin{document}
\title
{\bf Properties of superheavy nuclei with Z = 124}

\author{M. S. Mehta$^{1}$}
\author{Harvinder Kaur$^{1,2}$}
\author{Bharat Kumar$^{3}$}
\author{S. K. Patra$^{3}$}

\affiliation{\it $^{1}$Department of Applied Sciences, Punjab Technical University, 
Kapurthala 144 601, India}
\affiliation{\it $^{2}$Department of Physics, Rayat Bahra University, 
Mohali 140 104, India}
\affiliation{\it $^{3}$Institute of Physics, Sachivalaya Marg, 
Bhubaneswar 751 005, India}

\begin{abstract}

We employ Relativistic Mean Field (RMF) model with NL3 
parametrization to investigate the ground state properties of superheavy 
nucleus, Z = 124. The nuclei selected (from among complete isotopic series) 
for detailed investigation show that the nucleon density at the center 
is very low and therefore, these nuclei can be treated as semi-bubble 
nuclei. The considerable shell gap appears at neutron numbers 
N = 172, 184 and 198 showing the magicity corresponding to these numbers. 
The results are compared with the macro-microscopic Finite Range Droplet 
Model (FRDM) wherever possible.   
\end{abstract}
\maketitle
\footnotetext[1]{mehta\_iop@yahoo.co.uk}
\footnotetext[2]{patra@iopb.res.in}
\footnotetext[3]{bharat@iopb.res.in}
\section{Introduction}

The location of the center of 'island of stability' and hence the next magic 
number for proton beyond $^{208}$Pb (Z = 82, N = 126) in superheavy mass 
region is debated since the prediction of the existence of long-lived 
superheavy nuclei in sixties by \cite{sobi66,meld67,myer67,nils69,mosel69, 
seab69}. Since then a significant progress has been made in 
the discovery of superheavy nuclei \cite{hofm96,hofm00,kumar89}. 
Experimentally, the elements up to Z = 118 have been synthesized to-date, 
with half-lives varying from a few minutes to milliseconds 
\cite{hofm00}. Recently, the nuclei with Z = 104 - 118  
with mass number A = 266 - 294 have been detected at Dubna \cite{ogan99,
ogan00,yu01,ogan04,ogan07,eich07,ogan10,yu11} using “hot fusion” reactions 
with the neutron-rich $^{48}$Ca beam on actinides targets. 
These measurements show the increase in half-lives with  
in neutron number towards N = 184 give indication of stable center. 
In more detail, the cold fusion reactions involving a doubly magic spherical 
target and a deformed projectiles was used at GSI~\cite{hofm00,hof95,hofm95,
hofm96,hofm98,hofm99} to produce heavy elements upto Z = 110 - 112. 
At the  production time of Z = 112 nucleus at GSI the fusion cross-section was 
extremely small (1 pb), which led to the conclusion that reaching still
heavier elements will be very difficult. At this time, the emergence of hot
fusion reactions using $^{48}$Ca projectiles at Dubna has drastically changed
the situation and nuclei with Z = 114 - 118 were synthesized and also observed
their $\alpha$-decay chains. The element Z = 113 was first reported by  
Oganessian et al.~\cite{ogan04} and then using cold fusion reaction 
confirmed by Morita  et al.~\cite{morita04,morita12}.  \\

But theoretically, the studies of the shell structure of superheavy 
nuclei in different approaches show that the magic shells beyond 
the spherical double-magic number $^{208}$Pb (N = 126 and Z = 82), 
in superheavy mass region are isotope (combination of Z and N) as well 
as parameter dependent.
For example, recently, more microscopic calculations have 
predicted various other regions of stability, such as Z = 114, 
N = 184 \cite{rutz97}; Z = 120, N = 172 or 184 \cite{gupt97,patr99} 
and Z = 124 or 126, N = 184 \cite{cwio96,krup00,cwio05}.
In the fram-work of relativistic continuum Hartree-Bogoliubov theory, Zhang
et al.~\cite{zhan05} predicted Z = 120, 132 and 138 with neutron number 
N = 172, 184, 198, 228, 238 and 258 as the next nucleon shell gaps.
However, in experiments, the heaviest nucleus that 
could be studied so far is $^{254}$No (Z = 102, N = 152) \cite{herz06}. 
In an effort in this direction, using inductively coupled
plasma-sector field mass spectroscopy, Marinov {\it et al.} \cite{mari07}
have observed some neutron-deficient Th-isotopes in naturally
occurring thorium substances. 
The long-lived isomeric states,
with estimated half-lives T$_{1/2}\geqslant10^{8}$ y, have been identified
in the neutron-deficient $^{211,213,217,218}$Th isotopes, which are
associated with the super-deformed (SD) or hyper-deformed
(HD) states (minima) in potential energy surfaces (PES). 
In our earlier investigation \cite{patr09} of  Z = 122 isotopes
(N = 160 - 198), using relativistic mean field (RMF) and Skyrme 
Hartree Fock (SHF) models, we find the ground state solutions of 
some nuclei are super deformed and/or even hyper-deformed.
Of course, the SD ground state strcture of superheavy nuclei are reported
earlier  by Ren et al.~\cite{ren01}, within the theoretical 
framework of RMF calculations. 
Recently, Marinov {\it et al.} \cite{mari09} obtained a possible 
evidence for the existence of a long-lived superheavy nucleus with mass 
number A = 292 and atomic number Z = 122 or 124 in natural thorium. The 
half-life is again estimated to be the same as T$_{1/2}\geqslant 10^{8}$ y 
and the abundance is (1 - 10) $\times$ 10$^{−12}$ as compared 
to $^{232}$Th. This makes it interesting to make detailed investigation of 
the properties of nuclei in this mass region. 
\\

In extreme superheavy mass region, it is difficult to identify the nuclei 
by their $\alpha$-decay chains unless a proper combination of neutron and 
proton close shell are located.  Therefore, the identification of nuclei 
can be made through the comparison with theoretical calculations.
In the present investigation we calculate the bulk properties of 
Z = 124 nucleus within the framework of RMF model. 
Here, we choose NL3  parameter set~\cite{lala97} for isotopic chain with 
neutron number  N = 158 to N = 220, which encompasses the neutron 
numbers N = 172 and 184. Also, for the consistency of our 
results we calculate the similar quantities for isotopic chain of Z = 120 
nucleus.

\section{Formalism}

It has now been well established that the RMF models involving sigma, 
omega, rho and photon along with the self-interactions among various 
mesons, i.e., the effective field theory is very successful in 
explaining the structure of nuclei throughout the nuclear 
landscape~\cite{reinhard89,ring96,vret05,meng06,lian15}. 
The RMF model has been proved to be a very powerful tool to 
explain the properties of finite nuclei and infinite nuclear matter
\cite{mach89,patr91,patr01} for the last three decades.
We start with the modified relativistic Lagrangian density of 
$\sigma-\omega$ model \cite{sero86} for a 
nucleon-meson many-body system, which describes the nucleons as Dirac 
spinors interacting through the exchange of scalar mesons ($\sigma$), 
isoscalar vector mesons ($\omega$) and isovector mesons ($\rho$). The 
scalar mesons cause attraction and the vector mesons produce 
repulsion, whereas the charge protons generate electromagnetic interaction.

\begin{eqnarray}
{\cal L}&=&\overline{\psi_{i}}\{i\gamma^{\mu}
\partial_{\mu}-M\}\psi_{i}
+{1\over{2}}\partial^{\mu}\sigma\partial_{\mu}\sigma
-{1\over{2}}m_{\sigma}^{2}\sigma^{2}-{1\over{3}}g_{2}\sigma^{3}\nonumber\\
&-&{1\over{4}}g_{3}\sigma^{4}
-g_{s}\overline{\psi_{i}}\psi_{i}\sigma
-{1\over{4}}\Omega^{\mu\nu}\Omega_{\mu\nu}
+{1\over{2}}m_{w}^{2}V^{\mu}V_{\mu}\nonumber\\
&+&{1\over{4}}c_3(V_{\mu}V^{\mu})^2-g_{w}\overline\psi_{i}
\gamma^{\mu}\psi_{i}
V_{\mu}-{1\over{4}}\vec{B}^{\mu\nu}.\vec{B}_{\mu\nu}\nonumber\\
&+&{1\over{2}}m_{\rho}^{2}{\vec R^{\mu}}.{\vec{R}_{\mu}}
-g_{\rho}\overline\psi_{i}\gamma^{\mu}\vec{\tau}\psi_{i}
-{1\over{4}}F^{\mu\nu}F_{\mu\nu}\nonumber\\
&-&e\overline\psi_{i}
\gamma^{\mu}{\left(1-\tau_{3i}\right)\over{2}}\psi_{i}A_{\mu}.
\end{eqnarray}

The field for the $\sigma$-meson is denoted by $\sigma$, that for 
the $\omega$-meson by $V_{\mu}$ and for the isovector $\rho$-meson by 
$\vec R_{\mu}$. $A^{\mu}$ denotes the electromagnetic field. 
The $\psi_i$ are the Dirac spinors for the nucleons whose third component 
of isospin is denoted by $\tau_{3i}$. Here $g_{s}$, $g_{w}$, $g_{\rho}$ and
${e^{2}\over{4\pi}}={1\over{137}}$ are the coupling constants for 
$\sigma$, $\omega$, $\rho$ mesons and photon, respectively. $g_2$, $g_3$ 
and $c_3$ are the parameters for the nonlinear terms of $\sigma$- 
and $\omega$-mesons. M is the mass of the nucleon and $m_{\sigma}$, 
$m_{\omega}$ and $m_{\rho}$ are the masses of the $\sigma$, $\omega$ and
$\rho$-mesons, respectively. $\Omega^{\mu\nu}$, $\vec{B}^{\mu\nu}$ 
and $F^{\mu\nu}$ are the field tensors for the $V^{\mu}$, $\vec{R}^{\mu}$ 
and the photon fields, respectively
\cite{gamb90}.

From the relativistic Lagrangian, we get the field equations for the nucleons 
and mesons. These equations are solved by expanding the upper and lower 
components of Dirac spinors and the Boson fields in a deformed harmonic 
oscillator basis with an initial deformation. The set of coupled equations 
is solved numerically by a self-consistent iteration method. The center of 
mass motion is estimated by the usual harmonic oscillator formula 
$E_{c.m.}=\frac{3}{4}(41A^{-1/3})$ MeV. The quadrupole deformation parameter 
$\beta_2$ is evaluated from the resulting quadrupole moment \cite{gamb90} 
using the formula, 
\begin{eqnarray}
Q=Q_n+Q_p={\sqrt{{9\over{5\pi}}}}AR^2{\beta_{2}},
\end{eqnarray}
where $R=1.2A^{1/3}$ fm. The total binding energy of the system is,
\begin{eqnarray}
E_{total}=E_{part}+E_{\sigma}+E_{\omega}+E_{\rho}+E_c+E_{pair}+E_{c.m.},
\end{eqnarray}
where $E_{part}$ is the sum of the single-particle energies of 
the nucleons and $E_{\sigma}$, $E_{\omega}$, $E_{\rho}$, $E_c$ 
and $E_{pair}$ are the contributions of the mesons fields, the Coulomb 
field and the pairing energy, respectively.

For the open shell nuclei, 
the effect of pairing interactions is added in the BCS formalism.
We consider only T=1 channel of pairing correlation, i.e., pairing between 
proton-proton  and neutron-neutron. In such case, a nucleon of quantum 
state $|j, m_z\rangle$ pairs with another nucleons having same $I_z$ value 
with quantum state $|j,{-m_z}\rangle$, which is the time reversal partner 
of other. 
The RMF Lagrangian density only accommodates term like 
${\psi}^{\dagger}{\psi}$ (density) and no term of the form 
${\psi}^{\dagger}{\psi}^{\dagger}$ or $\psi\psi$. The inclusion 
of pairing correlation of the form $\psi \psi$ or 
${\psi}^{\dagger}{\psi}^{\dagger}$ violates the particle number 
conservation~\cite{patra93}. Thus, a constant gap BCS-type simple prescription 
is adopted in our calculations to take care of the pairing correlation
for open shell nuclei.  The general expression 
for pairing interaction to the total energy in terms of occupation 
probabilities $v_i^2$ and $u_i^2=1-v_i^2$ is written 
as~\cite{pres82,patra93}:
\begin{equation}
E_{pair}=-G\left[\sum_{i>0}u_{i}v_{i}\right]^2,
\end{equation}
with $G=$ pairing force constant. 
The variational approach with respect to the occupation number $v_i^2$ 
gives the BCS equation 
\cite{pres82}:
\begin{equation}
2\epsilon_iu_iv_i-\triangle(u_i^2-v_i^2)=0,
\label{eqn:bcs}
\end{equation}
with $\triangle=G\sum_{i>0}u_{i}v_{i}$. 

The densities with occupation number is defined as:
\begin{equation}
n_i=v_i^2=\frac{1}{2}\left[1-\frac{\epsilon_i-\lambda}{\sqrt{(\epsilon_i
-\lambda)^2+\triangle^2}}\right].
\end{equation}
For the pairing gap ($\triangle$) of proton and neutron is taken from 
the phenomenological formula of Madland and Nix \cite{madland}:
\begin{eqnarray}
\triangle_n=\frac{r}{N^{1/3}}exp(-sI-tI^{2})
\\
\triangle_p=\frac{r}{Z^{1/3}}exp(sI-tI^{2})
\end{eqnarray}
where, $I=(N-Z)/A$, $r=5.73$ MeV, $s=0.117$, and $t=7.96$.
 
The chemical potentials $\lambda_n$ and $\lambda_p$ are determined by the
particle numbers for neutrons and protons. The pairing energy of the 
nucleons using  equation (7) and (8) can be written as:
\begin{equation}
E_{pair}=-\triangle\sum_{i>0}u_{i}v_{i}.
\end{equation}

In constant pairing gap calculation, for a particular value of 
pairing gap $\triangle$ and force constant $G$, the pairing energy $E_{pair}$
diverges, if it is extended to an infinite configuration space.
In fact, in all realistic calculations with finite range forces,
the contribution of states of large momenta above the Fermi surface
(for a particular nucleus) to  $\triangle$ decreases with energy.
Therefore, the pairing window in all the equations are extended upto the level 
$|\epsilon_i-\lambda|\leq 2(41A^{-1/3})$ as a function of the single 
particle energy. 
The factor 2 has been determined so as to reproduce the pairing correlation 
energy for neutrons in $^{118}$Sn using Gogny force 
\cite{gamb90,patra93,dech80}.  
\section{Results and Discussions}

The superheavy nucleus Z = 124 with neutron number N = 158 - 220 are studied
for the investigation of ground state properties.  The results are compared 
with other models of previous works including the Finite Range Droplet 
Model (FRDM), as the experimental observations could not be made at such a 
high Z region so far. In numerical calculations, the number of oscillator 
shell for Fermions and Bosons N$_{F}$ = N$_{B}$ = 20 are used to evaluate 
the physical observables with the pairing gaps of eqns. (7) and (8) in the BCS
pairing scheme.

\subsection{Binding Energy}

\begin{figure}[h]
\includegraphics[width=1.\columnwidth]{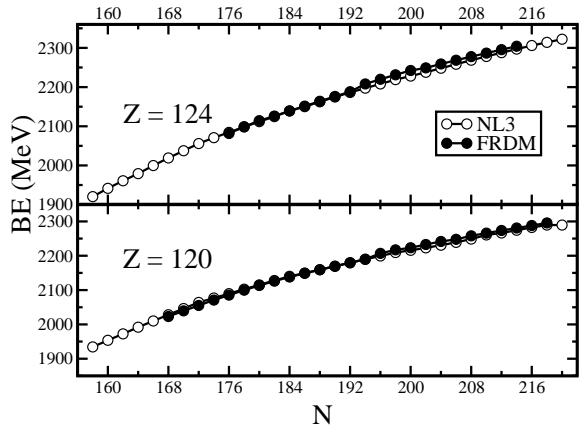}
\caption{\label{fig1} The binding energy of isotopic series of 
nuclei Z = 120, 124 nuclei with NL3 parameter set.}
\label{bea1}
\end{figure}

The binding energy of the isotopic chain of Z = 124 is calculated for 
mass number A = 282-384. Since there is no experimental observation 
for such a large Z number so far, therefore, the only comparison 
can be made with theoretical models such as macroscopic-microscopic model. 
We compare our calculations with finite range droplet model 
(FRDM) \cite{moel97}. Here in upper panel of Fig.~\ref{bea1}, we compare the 
results (binding energy) with available FRDM results, which 
seem to be in good agreement. A small difference in binding energy 
at N = 198 region can be seen in the upper panel of Fig.~\ref{bea1}.
For example, the RMF results of binding energy and quadrupole deformation
parameter for $^{334}$124 nucleus are 2284.71 MeV and $\beta_{2}$ = 0.128,
whereas the FRDM calculations are 2286.75 MeV and $\beta_{2}$ = 0.335, 
respectively. Similarly for $^{312}$124, the RMF binding energy is 2166.19
MeV and FRDM value is 2163.84 MeV with a discrepancy of 2.35 MeV. In Z = 124
isotopes, we get a maximum difference in binding energy is 7.43 MeV for 
$^{320}$124 nucleus, which is about 0.3 \% discrepancy. In general, the 
difference in binding 
energy with FRDM and RMF is $\sim$3 - 4 MeV, which is reasonable in the order
of two thousands magnitude.
   For consistency of our results, we also  calculate the binding energy 
of the isotopic chain of Z = 120 and is displayed at the lower panel of 
Fig.~\ref{bea1}. In this case the difference in binding energy is very small.
For example, the RMF result of BE and $\beta_{2}$ are 2026.51 MeV and 
 $\beta_{2}$ = -0.049 compared to the FRDM,binding enrgy BE = 2023.06 MeV and 
$\beta_{2}$ = -0.104 for  $^{288}$120. Similarly, maximum discrepancy 
between RMF and FRDM bunding energy is $\sim$5.83 MeV for $^{320}$120.

\subsection{Separation Energy}
The magic numbers in nuclei are characterized by the large shell gap in 
single particle energy levels. This means the nucleon in lower level has 
comparatively large value of energy than that in higher level giving rise to the 
more stability. 
The extra stability corresponding to certain numbers can be estimated from 
the sudden fall in the neutron separation energy. The separation energy is 
calculated by the difference in binding energies of two isotopes using relation:

\begin{equation}
S_{2n}(N,Z) = BE(N,Z)-BE(N-2, Z)\nonumber
\end{equation}
 
\begin{figure}[h]
\includegraphics[width=1.\columnwidth]{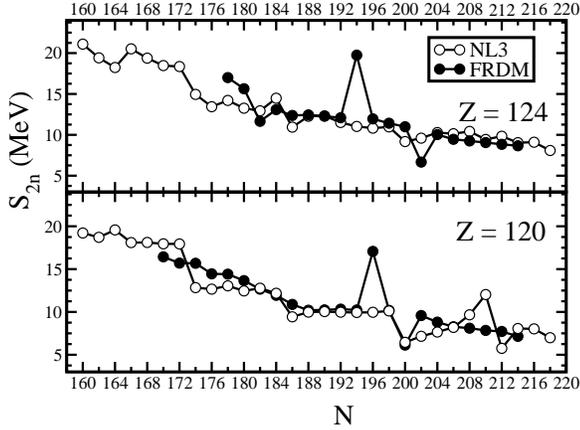}
\caption{\label{fig1} The two neutron separation energy as a function of 
neutron for series of Z = 120 and 124 nuclei.}
\label{bea2}
\end{figure}

The two neutron separation energy (S$_{2n}$) for the isotopic series 
of nuclei Z=124 and 120 ($^{282-344}$124 and $^{278-340}$120) is shown 
in Fig.~\ref{bea2}. The sudden 
fall in separation energy at N = 172, 184 and 198 can clearly be seen 
in both the cases confirming the magic character 
\cite{rutz97,gupt97,patr99,meht02} predicted in earlier studies. 
Although, N = 172 is not that much pronounced in our earlier  
investigation of odd nuclei \cite{meht02}, here the magicity at N = 172 
increases as we move to extreme of superheavy mass region \cite{sald09}.
Contrary to some earlier literature, there is no signature of sudden change in 
separation energy at deformed magic number N = 162 \cite{rutz97} in 
the present calculations.  
The decrease in energy at N = 172 
and 184 is $\sim$ 3.5 MeV whereas $\sim$2 MeV at N = 198, for Z = 124 
nuclei. 
In case of Z = 120 isotopes the decrease in energy is $\sim$ 5.0 MeV 
at N = 172 and $\sim$ 3.0 MeV and 3.5 MeV at N =184 and 198 respectively. 
Such decrease at N = 198 in our calculation is nearly same as in FRDM value.  
However, in FRDM the sudden decrease in 
separation energy appears at N = 180 and 200 for Z = 124. Except the values 
at these numbers, in general all other energies from our present calculations 
are in good agreement (within $\sim 2$ MeV accuracy) with macro-microscopic 
calculations (FRDM). We observed a couple of abnormal increase in $S_{2n}$ at 
(N=194, Z=124) and (N=196, Z=120), which are not seen in the present RMF 
calculations.

\subsection{Quadrupole Deformation Parameter}

\begin{figure}
\includegraphics[width=1.\columnwidth]{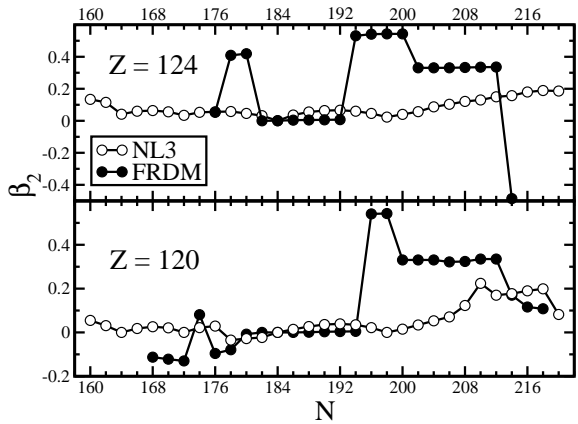}
\caption{\label{fig1} The quadrupole deformation parameter 
$\beta_2$ for the isotopic series of Z = 120, 124 nuclei.}
\label{bea3}
\end{figure}

The quadrupole deformation parameter $\beta_2$ gives the shape of nuclei 
in ground state. The value of $\beta_2$ is positive, negative and zero for 
prolate, oblate and spherical respectively. In our calculation 
shown in Fig.~\ref{bea3}. except for few nuclei all the isotopes of Z = 124 are 
either spherical or near spherical. The results compared with FRDM 
\cite{moel97} agree for nuclei having N = 176, 182 - 192 as shown in 
the upper panel of Fig.~\ref{bea3}. At N = 176 and N = 184 the nuclei are 
complete  spherical. There is the least agreement beyond N = 196 for Z = 124. 
From the  figure it is clear that NL3 parameter set predicts the deformation 
parameter $\beta_2$ very close to FRDM at middle mass region i.e., from 
neutron number N = 176 to 192. 

\subsection{Q$_{\alpha}$ Energy and Half-Life (T$_{\alpha}$)}

The superheavy nuclei along to the $\beta$-stability line are known to 
be $\alpha$-emitter. The $\alpha$-decay half-life of the nucleus showing 
shell closure is believed to be comparatively larger than the neighboring 
nuclei. 
Thus, to confirm the magic number corresponding to a particular neutron 
number N, it is beneficial to calculate half-life of $\alpha$-decay. 
The investigation of $\alpha$-decay of nuclei gives information 
about their degree of stability and possibility of 
existence in nature. 
Here we take the nucleus $^{296}$124 
(Z = 124 and N = 172) for the calculation of $\alpha$-decay 
energy \cite{patr97}. 

\begin{figure}[b]
\includegraphics[width=1.\columnwidth]{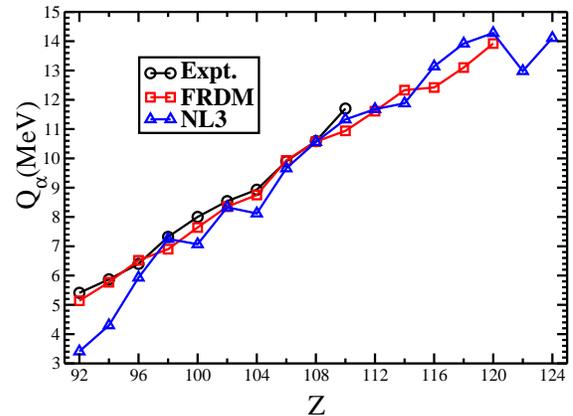}
\caption{\label{fig1} (Color online) The $\alpha$-decay ($Q_\alpha$-energy) 
chain from Z = 124 to Z = 92.}
\label{bea4}
\end{figure}

The Q$_{\alpha}$-energy and half life (T$_{\alpha}$) are compared 
with available experimental data as shown in Table 1. 
The Q$_{\alpha}$-energy is calculated using the following equation: 

\begin{equation}
Q_{\alpha}(N,Z) = BE(N,Z)-BE(N-2, Z-2)- BE(2,2)\nonumber.
\end{equation}
\\
In the equation, BE(N,Z) is binding energy of the parent nucleus having 
N neutrons and Z protons, and BE(N - 2, Z - 2) is the binding energy of 
daughter nucleus after emission of an $\alpha$-particle (BE(2,2)). The 
binding energy of $\alpha$-particle ($^{4}$He) is 28.296 MeV. 
The Q$_{\alpha}$ energy 
values are in good agreement with experimental data \cite{audi03} 
as well as FRDM \cite{moel97} as shown in Table. 
The decay chain is also plotted in Fig.~\ref{bea4} which shows good agreement 
with experiments as well as FRDM calculations. 
The half-life log$_{10}$ T$_\alpha$(s) values are estimated using the 
phenomenological formula \cite{viol66};

\begin{equation}
log_{10} T_{\alpha}(s) = \frac{aZ - b}{\sqrt{Q_\alpha}} 
- (cZ+d)-h_log,\nonumber 
\end{equation} 
where Z is atomic number of parent nucleus, and the other 
parameters are; a = 1.66175, b = 8.5166, c = 0.20228, and 
d = 33.9069. The values of the parameters are used from 
Sobiczewski {\it et al.} \cite{sobi89}. The hindrance 
($h_{log}$) caused by odd number of protons and/or 
neutrons is zero here.

\begin{widetext}

\renewcommand{\arraystretch}{1.3}
\begin{table}
\caption{ The Q$_\alpha$ and T$_\alpha$ calculated using NL3
parameter set in RMF. The results are compared with finite range
droplet model (FRDM)\cite{moel97} as well as the available
experimental data\cite{audi03}.The binding energy
is in MeV and half life is in seconds.}

\vspace{0.10 in}

\begin{tabular}{ccccccccccccccccc}
\hline
&\multicolumn{9}{c}{RMF(NL3)}
&\multicolumn{4}{c}{FRDM}
&\multicolumn{3}{c}{Expt.}\\
\hline
\multicolumn{1}{c}{A}
&\multicolumn{2}{c}{Z}
&\multicolumn{2}{c}{BE}
&\multicolumn{2}{c}{Q$_\alpha$}
&\multicolumn{2}{c}{T$_\alpha$}
&\multicolumn{2}{c}{BE}
&\multicolumn{2}{c}{Q$_\alpha$}
&\multicolumn{1}{c}{T$_\alpha$}
&\multicolumn{1}{c}{BE}
&\multicolumn{1}{c}{Q$_\alpha$}
&\multicolumn{1}{c}{T$_\alpha$}\\
\hline
296&&124&&2056.01&&14.11&&10$^{-6.41}$&&&&&&&&\\
292&&122&&2041.83&&12.98&&10$^{-4.68}$&&&&&\\
288&&120&&2026.51&&14.28&&10$^{-7.66}$&&2023.06&&13.92&10$^{-7.02}$&&\\
284&&118&&2012.49&&13.92&&10$^{-7.50}$&&2008.69&&13.10&10$^{-5.95}$&&\\
280&&116&&1998.12&&13.14&&10$^{-6.54}$&&1993.49&&12.42&10$^{-5.10}$&&\\
276&&114&&1982.96&&11.88&&10.$^{-4.48}$&&1977.62&&12.33&10$^{-5.44}$&&\\
272&&112&&1966.55&&11.68&&10$^{-4.60}$&&1961.66&&11.61&10$^{-4.45}$&&\\
268&&110&&1949.93&&11.33&&10$^{-4.38}$&&1944.97&&10.94&10$^{-3.47}$
&1943.53&11.7&10$^{-5.2}$\\
264&&108&&1932.96&&10.56&&10$^{-3.14}$&&1927.62&&10.57&10$^{-3.18}$
&1926.67&10.59&10$^{-3.2}$\\
260&&106&&1915.22&&9.66&&10$^{-1.42}$&&1909.90&&9.93&10$^{-2.15}$
&1909.06&9.90&10$^{-2.07}$\\
256&&104&&1896.59&&8.12&&10$^{2.73}$&&1891.53&&8.75&10$^{0.59}$
&1890.56&8.93&10$^{0.05}$\\
252&&102&&1876.41&&8.33&&10$^{1.25}$&&1871.98&&8.35&10$^{1.19}$
&1871.35&8.54&10$^{0.52}$\\
248&&100&&1856.44&&7.07&&10.$^{5.15}$&&1852.03&&7.64&10$^{2.91}$
&1851.57&8.0&10$^{1.60}$\\
244&&98&&1835.21&&7.25&&10$^{3.57}$&&1831.38&&6.90&10$^{5.01}$
&1831.22&7.32&10$^{3.30}$\\
240&&96&&1814.17&&5.93&&10$^{8.68}$&&1809.98&&6.52&10$^{5.81}$
&1810.28&6.40&10$^{6.36}$\\
236&&94&&1791.81&&4.30&&10$^{18.29}$&&1788.21&&5.77&10$^{8.54}$
&1788.41&5.87&10$^{8.03}$\\
232&&92&&1767.81&&3.41&&10$^{25.66}$&&1765.695&&5.14&10$^{11.18}$
&1765.98&5.41&10$^{9.50}$\\
\hline
\end{tabular}
\label{table4}
\end{table}
\end{widetext}

\subsection{Density Distribution}
\begin{figure}
\includegraphics[width=1.\columnwidth]{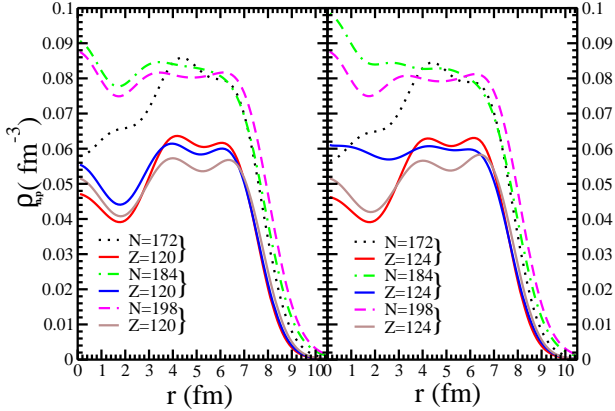}
\caption{\label{fig1} (Color online) The density of selected isotopes of 
Z = 120, 124 nuclei with NL3 parameter set.}
\label{bea5}
\end{figure}

The neutron and proton density distributions for Z = 124 and 120 
nuclei are plotted in Fig.~\ref{bea5}. The nuclei with N = 172, 184 and 198 
are taken as representative cases for detailed investigation of the 
internal structure.
The charge distribution of both Z = 120 and 124 show that the center part 
of nuclei have very low density indicating a hollow inside. 

\begin{figure}
\includegraphics[width=1.\columnwidth]{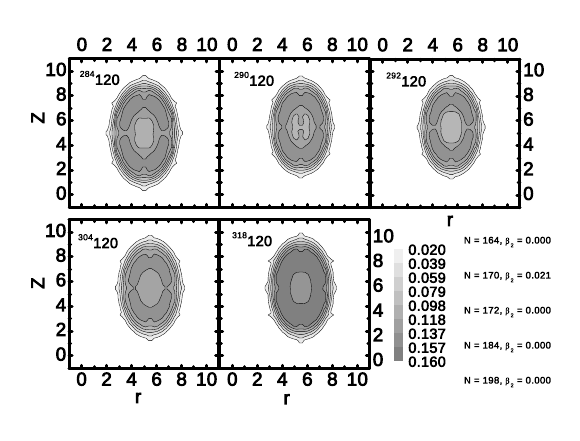}
\caption{Two-dimensional density contours for nuclei 
$^{284,290,292,304,318}$120 shown using  NL3 parameter set.}
\label{bea6}
\end{figure}

\begin{figure}
\includegraphics[width=1.\columnwidth]{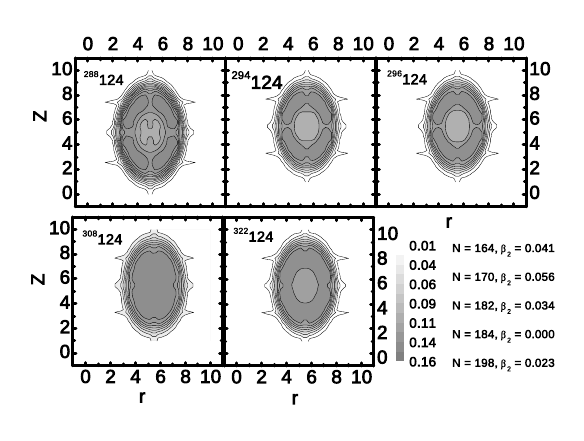}
\caption{\label{fig2} Same as Fig. 6 but for nuclei 
 $^{288,294,296,308,322}$124.}
\label{bea7}
\end{figure}

\begin{figure}
\includegraphics[width=1.\columnwidth]{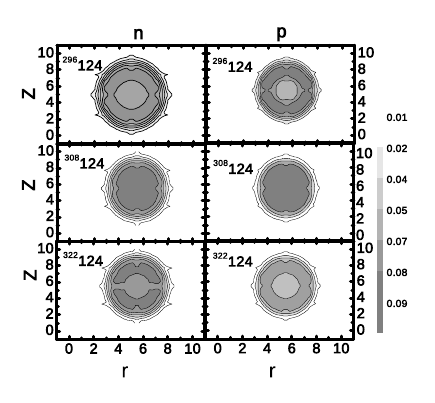}
\caption{\label{fig3} The neutrons and protons density distribution for
nuclei $^{296,308,322}$124.}.
\label{bea8}
\end{figure}

In order to have an insight into the arrangement of nucleons we plot the 
two dimensional contours for some selected nuclei. 
The density contours for $^{284,290,292,304,318}$120 
and $^{288,294,296,308,322}$124 nuclei are shown 
in Figs.~\ref{bea6},~\ref{bea7}, and ~\ref{bea8}. In general, it is clear 
from the figures that the central region in all nuclei except $^{308}$124 
nucleus have considerably low density. In case of isotopes of Z = 120, as 
shown in Fig.~\ref{bea6}, N = 170 nucleus is slightly deformed 
($\beta_2$ = 0.021) and all other(N = 164, 170, 184 and 198) are spherical 
in their ground state. The nuclei with Z$\geqslant$120 have large number of 
protons and hence considerable Coulomb repulsion among protons. The strong 
repulsion changes the entire distribution of nucleons. The doubly magic 
nucleus $^{292}$120 is largely studied previously 
\cite{bend99,dech99,pei05} and is predicted to be semi-bubble.
In the present calculations using RMF(NL3), semi-bubble structure of 
these nuclei can be clearly seen in Fig.~\ref{bea6}.
The hollow region at the center is spread over the 
radius of 1 - 2 fm. This may suggest that these nuclei might be a 
fullerene type structure 
consisting of 60 $\alpha$-particles and a binding neutron per alpha and/ or 
few neutron clusters. The clusters of some heavier nuclei might be possible.
The density distribution of $^{288,294,296,308,322}$124 nuclei is 
shown in Fig.~\ref{bea7}. In this case the density of nucleus N = 184 is 
more at the central region while all other nuclei studied here are showing 
bubble type structure. The low density region extends up to $\sim$ 2 fm. 
The nuclei with N = 164, 170, 172 and 198 are near spherical 
($\beta_2$ = 0.041, 0.056, 0.034 and 0.023 respectively) whereas N = 184 
is spherical in shape.
In order to give further insight into the arrangement of nucleons, 
we plot the density distribution of neutron and proton separately 
(Fig.~\ref{bea8}). It is clear from the figure that both neutrons as well as 
protons are shifted from the central region except for N = 184 nucleus. \\

\section{Conclusion}
In the present work we use RMF(NL3) model to explore the structure of 
superheavy nucleus Z = 124. The results of our calculations are 
compared with macro-microscopic FRDM prediction. We calculate binding 
energy, quadrupole deformation parameter ($\beta_2$), two neutron 
separation energy (S$_{2n}$), and decay half-life (T$_{1/2}$) 
for the isotopic series of Z = 124 and for the consistence of our results 
we calculate the same quantities for Z = 120 nucleus. The quadrupole 
deformation parameter at heavier side of series show more deviation from 
FRDM values. The two neutron separation energy shows the sudden fall in 
energy at neutron numbers N = 172, 184, and 198 indicating the magic structure. 
The $\alpha$-decay energy and half-life are also calculated and compared 
with the experiments and FRDM results which seem to be in good agreement. 
The density profile of the selected nuclei shown that the depression in 
the density at the central region of the nuclei with the exception 
of $^{308}$124. This nucleus is the only candidate which does not 
show the depression at the center. Finally, this theoretical investigation 
of ground state properties of Z = 124 nuclei may be helpful for an 
experimental exploration to locate the ``island of stability'' which is 
expected to be existed in the large Z superheavy region.

\end{document}